\documentclass[prl,twocolumn,aps,showkeys,showpacs,longbibliography]{revtex4-1}

\usepackage{graphicx} 
\usepackage{amsmath} 
\usepackage{amssymb}
\usepackage{color}

\newcommand{\vd}[1]{\mathrm{d}\vec{#1}}

\begin{document}

\title{Efficient high-dimensional entanglement imaging with a
  compressive sensing, double-pixel camera}

\author{Gregory A. Howland}
\email{ghowland@pas.rochester.edu} 
\author{John C. Howell}

\affiliation{Department of Physics and Astronomy, University of
  Rochester, Rochester, New York, 14627, USA}

\date{\today}

\begin{abstract}
  We implement a double-pixel, compressive sensing camera to
  efficiently characterize, at high resolution, the spatially
  entangled fields produced by spontaneous parametric
  downconversion. This technique leverages sparsity in spatial
  correlations between entangled photons to improve acquisition times
  over raster-scanning by a scaling factor up to $n^2/\log(n)$ for
  $n$-dimensional images. We image at resolutions up to $1024$
  dimensions per detector and demonstrate a channel capacity of $8.4$
  bits per photon. By comparing the entangled photons' classical
  mutual information in conjugate bases, we violate an entropic
  Einstein-Podolsky-Rosen separability criterion for all measured
  resolutions. More broadly, our result indicates compressive sensing
  can be especially effective for higher-order measurements on
  correlated systems.
\end{abstract}

\pacs{03.65.Ud, 03.65.Wj, 03.67.Mn, 89.70.Cf, 89.70.Eg}
\keywords{Quantum Information, Computational Physics}
\maketitle

\section{Introduction}
Spatially entangled biphotons, such as those generated by spontaneous
parametric downconversion (SPDC), exhibit strong
Einstein-Podolsky-Rosen (EPR) type correlations \cite{einstein:1935}
in the transverse position and transverse momentum degrees of freedom
\cite{howell:2004}. Because these variables are continuous, the
entanglement can be very high-dimensional with a typical Schmidt
number greatly exceeding $1000$ \cite{pors:2011:thesis}. This provides
high information density which can be leveraged to increase channel
capacity and security for quantum key distribution \cite{ekert:1991,
  walborn:2006, walborn:2008} and dense coding
\cite{bennett:1992,braunstein:2000}. Other applications include ghost
imaging \cite{pittman:1995,abouraddy:2001}, quantum computing
\cite{tasca:2011}, and quantum teleportation \cite{walborn:2007}.

Experimentally characterizing the SPDC state is unfortunately
difficult due to weak sources and low resolution detectors. Spatial
entanglement is traditionally imaged by jointly raster-scanning
photon-counting avalanche photodiodes (APDs) to measure spatial
correlations. This scales extremely poorly with increasing detector
resolution. With a biphoton flux of $4,000$ coincident detections per
second, it would take $55$ days to jointly scan a $24\times 24$ pixel
region for a signal-to-noise ratio (SNR) of $10$. For $32 \times 32$
pixels, it would take $310$ days (see Eq. \ref{eq:raster}).

Other approaches have been tried with mixed success. Intensified CCD
cameras can measure the Schmidt number \cite{pires:2009}, but do not
detect single-photon correlations rendering them ineffective for most
quantum applications. Arrays of photon counting APDs could replace
CCDs, but they are currently low resolution, noisy and resource
intensive; especially since each pixel pair must be individually
correlated \cite{coffey:2011,albota:2002,itzler:2010}. A recent,
promising result averages intensity correlations over many images from
a single-photon sensitive electron-multiplying CCD, reporting $2500$
modes \cite{edgar:2012}. This technique is limited to a $30$ ms
exposure time (APDs are sub-ns) and is noisier than using APDs because
it does not isolate individual coincident detections.

In ref \cite{dixon:2012}, Dixon \emph{et. al.} reduce the number of
measurements required for a raster-scan by only measuring in an area
of interest where correlations are expected, reporting a channel
capacity of $7$ bits per photon. While not a true full-field
measurement, they highlight a critical feature of the SPDC field. In
\emph{both} position and momentum representations, the distribution of
correlations between pairs of detector pixels is very sparse despite
dense (not sparse) single-particle distributions. Applying ideas from
the field of compressive sensing, we exploit prior knowledge of
this sparsity to beat the ``curse of dimensionality''
\cite{bellman:2003} and efficiently characterize the full biphoton
field without raster scanning.

In this article, we implement a compressive sensing, photon-counting
double-pixel camera that efficiently images single-photon SPDC
correlations in the near- and far-field at resolutions up to $32
\times 32 = 1024$ dimensions per detector. At $32 \times 32$
resolution, the measurement time is reduced from $310$ days for raster
scanning to around $8$ hours. We perform an entropic characterization
showing channel capacities of up to $8.4$ bits per photon, equivalent
to $337$ independent, identically distributed modes. Sums of channel
capacities in conjugate bases violate a EPR steering bound
\cite{schneel:2012} by up to $6.6$ bits.

\section{Theory}
\subsection{Compressive Sensing}
Compressive sensing is a technique that employs optimization to
measure a sparsely represented $N$-dimensional signal from $M<N$
incoherent measurements
\cite{donoho:2006,candes:2007,baraniuk:2007,candes:2008}. The approach
is so-named because the signal is effectively compressed during
measurement. Though sparsity is assumed, it is not known prior to
measurement which elements contain appreciable amplitude. Compressive
sensing must determine both which elements are significant and find
their values.

To detect a sparsely represented $N$-dimensional signal vector $X$, we
measure a series of $M<N$ values $Y$ by multiplying $X$ by an $M
\times N$ sensing matrix $A$ such that $\bm{}$
\begin{equation}
  Y = \bm{AX} + \bm{\Gamma},
  \label{eq:meas}
\end{equation}
where $\Gamma$ is a noise vector.

Because $M<N$, this system is undetermined; a given $Y$ does not
specify a unique $X$. The correct $X$ is recovered by
minimizing a regularized least squares objective function
\begin{equation}
  \min_X \frac{1}{2}||Y-AX||^{2}_{2}+\tau g(X),
  \label{eq:objective}
\end{equation}
where for example $||\Omega||^{2}_{2}$ is the $\ell_2$ norm of
$\Omega$ and $\tau$ is a scaling constant. The function $g(X)$ is a
regularization promoting sparsity. Common $g(X)$ include $X$'s
$\ell_1$ norm, assuming the signal is sparse, and $X$'s total
variation, assuming the signal's gradient is sparse
\cite{candes:2005}. $A$ must be incoherent with the basis of interest,
with the surprising and non-intuitive result that a random, binary
sensing matrix works well. Given sufficiently large $M$, the recovered
$X$ approaches the exact signal with high probability
\cite{candes:2006}. For a $k-$sparse signal, the required $M$ scales
as $M \propto k \log (N/k)$.

Incoherent, random sampling is particularly beneficial for low-light
measurements as each measurement receives on average half the total
photon flux $\Phi/2$, as apposed to $\Phi/n$ for a raster
scan. Compressive sensing is now beginning to be used for quantum
applications such as state tomography \cite{gross:2010}. Shabani
et. al, for example, perform a tomography on a two qubit photonic gate
for polarization entangled photons \cite{shabani:2011}. CS has also
been used to with spatially correlated light for ghost imaging
\cite{katz:2009, zerom:2011}. It is important to note that for ghost
imaging, CS is not required to recover the full two-particle
probability distribution as in entanglement characterization.

The quintessential compressed sensing example is the single-pixel
camera \cite{wakin:2006, duarte:2008}. An object is imaged onto a
Digital Micromirror device (DMD), a 2D binary array of
individually-addressable mirrors that reflect light either to a single
detector or a dump. Rows of the sensing matrix $A$ consist of random,
binary patterns placed sequentially on the DMD. For an $N$-dimensional
image, minimizing Eq. \ref{eq:objective} recovers images using as few
as $M = 0.02N$ measurements.

\subsection{Compressive Sensing for Measuring Correlations}
The single-pixel camera concept naturally adapts to imaging
correlations by adding a second detector. Consider placing separate
DMDs in the near-field or far-field of the SPDC signal and idler
modes, where ``on'' pixels are redirected to photon counting
modules. The signal of interest is
\begin{align}
  p_{x}(u,v) & = \int \limits_u \vd{x_s} \int \limits_v \vd{x_i}
  |f(\vec{x_s},\vec{x_i})|^2 \\ p_{k}(u,v) & = \int \limits_u \vd{k_s}
  \int \limits_v \vd{k_i} |f(\vec{k_s},\vec{k_i})|^2,
  \label{eq:p}
\end{align}
where $p(u,v)$ represents the probability of a coincident detection
between the $u^{\text{th}}$ pixel on the signal DMD and $v^{\text{th}}$ pixel on the
idler DMD. The functions $f(\vec{x_s},\vec{x_i})$ and
$f(\vec{k_s},\vec{k_i})$ are approximate position and momentum
wavefunctions for the biphoton
\begin{alignat}{3}
  \psi(\vec{x_s},\vec{x_i}) = & \mathcal{N} \exp \left(-\frac{(\vec{x_s}-\vec{x_i})^2}{4 \sigma_c^2}\right)\exp \left(-\frac{(\vec{x_s}+\vec{x_i})^2}{16 \sigma_p^2}\right)\nonumber \\
  \psi(\vec{k_s},\vec{k_i}) = & (4\sigma_p\sigma_c)^2\mathcal{N} \exp (-\sigma_c^2(\vec{k_s}-\vec{k_i})^2) \nonumber \\
  & \times \exp (-4\sigma_p^2(\vec{k_s}+\vec{k_i})^2).
  \label{eq:wavefunction}
\end{alignat}
Subscripts $s$ and $i$ refer to signal and idler photons respectively,
$\sigma_p$ and $\sigma_c$ are the pump and correlation widths, and
$\mathcal{N}$ is a normalizing constant. $\bm{X}$ of
Eq. \ref{eq:objective} is simply a one-dimensional reshaping of $p_x$
or $p_k$.

Like the single-pixel camera, a series of random patterns are placed
on the DMDs to form rows of $A$. For each pair of patterns,
correlations between the signal and idler photons form the measurement
vector $Y$. Minimization of Eq. \ref{eq:objective} recovers $p(u,v)$.

While a fully random $\bm{A}$ is preferred, the DMDs only act on their
respective signal or idler subspace, which prevents arbitrary
$\bm{A}$. Rows of $\bm{A}$ are therefore outer products of rows of
single-particle sensing sensing matrices $\mathbf{a}$ and $\mathbf{b}$
\begin{equation} 
A = \begin{pmatrix}
      \mathbf{a}_1 \otimes \mathbf{b}_1\\
      \mathbf{a}_2\otimes \mathbf{b}_2\\
      \vdots\\
      \mathbf{a}_m \otimes \mathbf{b}_m
    \end{pmatrix},
\label{eq:kronmat}
\end{equation}
where rows of $\mathbf{a}$ represent random patterns placed on the
signal DMD and rows of $\mathbf{b}$ represent random patterns placed
on the idler DMD. To make signal and idler photons distinguishable,
$\mathbf{a}$ and $\mathbf{b}$ are not the same. The validity of
Kronecker-type sensing matrices has been established and is of current
interest in the CS community as attention shifts to higher dimensional
signals \cite{duarte:2010, duarte:2012}. The measurement vector $Y$ is
obtained by counting coincident detections for the series of DMD
configurations given by $A$.

A variety of reconstruction algorithms exist for
Eq. \ref{eq:objective}, with their computational complexity dominated
by repeatedly calculating $\bm{AX}$ and $\bm{A^TY}$
\cite{li:2010}. This is especially unwieldy for correlation
measurements as the size of $A$ is $M\times n^2$ for $n-$pixel DMDs.
Using properties for Kronecker products \cite{horn:1994}, these can be
more efficiently computed by
\begin{align}
  \bm{AX}  & = \textbf{diag}(\mathbf{b}\, \textbf{sq}(\bm{X})\, \mathbf{a}^T)\\
  \bm{A^TY} & = \textbf{vec}(\mathbf{b}^T\, \textbf{od}\,(\bm{Y})
  \mathbf{a}),
\end{align}
where $\textbf{sq}$ and $\textbf{vec}$ reshape a vector to a square
matrix and vice-versa; $\textbf{diag}$ forms a vector from the
diagonal elements of a square matrix; and $\textbf{od}$ forms a square
matrix placing the operand vector on its diagonal.

\subsection{Comparison to Raster Scanning}
The compressive approach finds the joint probability distribution
orders of magnitude faster than raster scanning through two key
improvements. The first is simply the reduction in the number of
measurements. To jointly raster scan an $n$-pixel space requires $n^2$
measurements. For a compressive measurement, sparsity is approximately
$n$ with dimensionality $n^2$, so only $M \propto n\log(n)$
measurements are required. In practice, we found excellent results
when $M$ was only three percent of $n^2$.

The second advantage of compressive measurements is that they more
efficiently use available flux. For the raster scan, the total flux is
distributed over at best $n$ pairs of pixels in the case of perfect
correlations. Conversely, the average flux per incoherent compressive
measurement is independent of $n$, with each measurement receiving on
average $1/4$ the total flux. To maintain constant SNR
(photons/measurement) with increasing $n$, total measurement time
therefore scales as $n^3$ for raster scanning. Given a photon flux of
$\Phi$ photons per second, the measurement time for a desired SNR is
\begin{equation}
t = n^2t_{\text{meas}} = \frac{n^3 \text{SNR}^2}{\Phi},
\label{eq:raster}
\end{equation}
where $t_{\text{meas}}$ is the time per measurement.

For incoherent, compressive measurements, acquisition time scales as
$n\log(n)$. The compressive improvement therefore scales as
$n^2/\log(n)$. For $n = 32 \times 32 = 1024$, this is of order $10^5$.

This scaling factor somewhat optimistically assumes the reconstruction
process yields an accurate result despite a noisy
signal. Unfortunately, propagation of uncertainty through the
reconstruction process remains a difficult problem, especially for
non-ideal, real world systems \cite{willett:2011}. There has been much
recent theoretical work on the topic for Gaussian
\cite{montanari:2011,wu:2012,reeves:2012} and Poissonian noise
\cite{willett:2009,harmany:2009}. These results tend to require ideal
sensing matrices or more complicated formulations to give provable
performance bounds. As such, their findings are difficult to directly,
quantitatively apply to experiment. However, they do reveal pertinent
features that indicate CS can perform extremely well in the presence
of noise.

A well known characteristic of CS is a rapid phase change from poor to
good quality reconstructions \cite{ganguli:2010}. This phase change is
often discussed as a function of increasing $m$, with the boundary $m
\propto k\log(n/k)$. A similar phase transition occurs as a function
of the noise level; in our case, this is the number of photons per
measurement. For some cases, these two are even linked
\cite{montanari:2011}. A practical compressive measurement simply
requires large enough $m$ and photon flux $\Phi$ to be in the space of
good reconstructions. Fortunately, simply obtaining a recognizable
reconstruction generally indicates the measurement conditions exceed
this threshold.

Unlike a direct measurement, the information obtained by a series of
$y$ compressive measurements is contained in their deviation from the
average value $\bar{y}$. In the presence of noise, these deviations
must exceed the noise level. Assuming Poissonian shot noise, good
reconstructions require $\text{std}(y) \ge \beta \sqrt{\bar{y}}$, where
$\text{std}(y)$ is the standard deviation of the measurement vector
and $\beta$ is a positive constant greater than one.

\begin{figure}[b]
\includegraphics[scale=0.35 ]{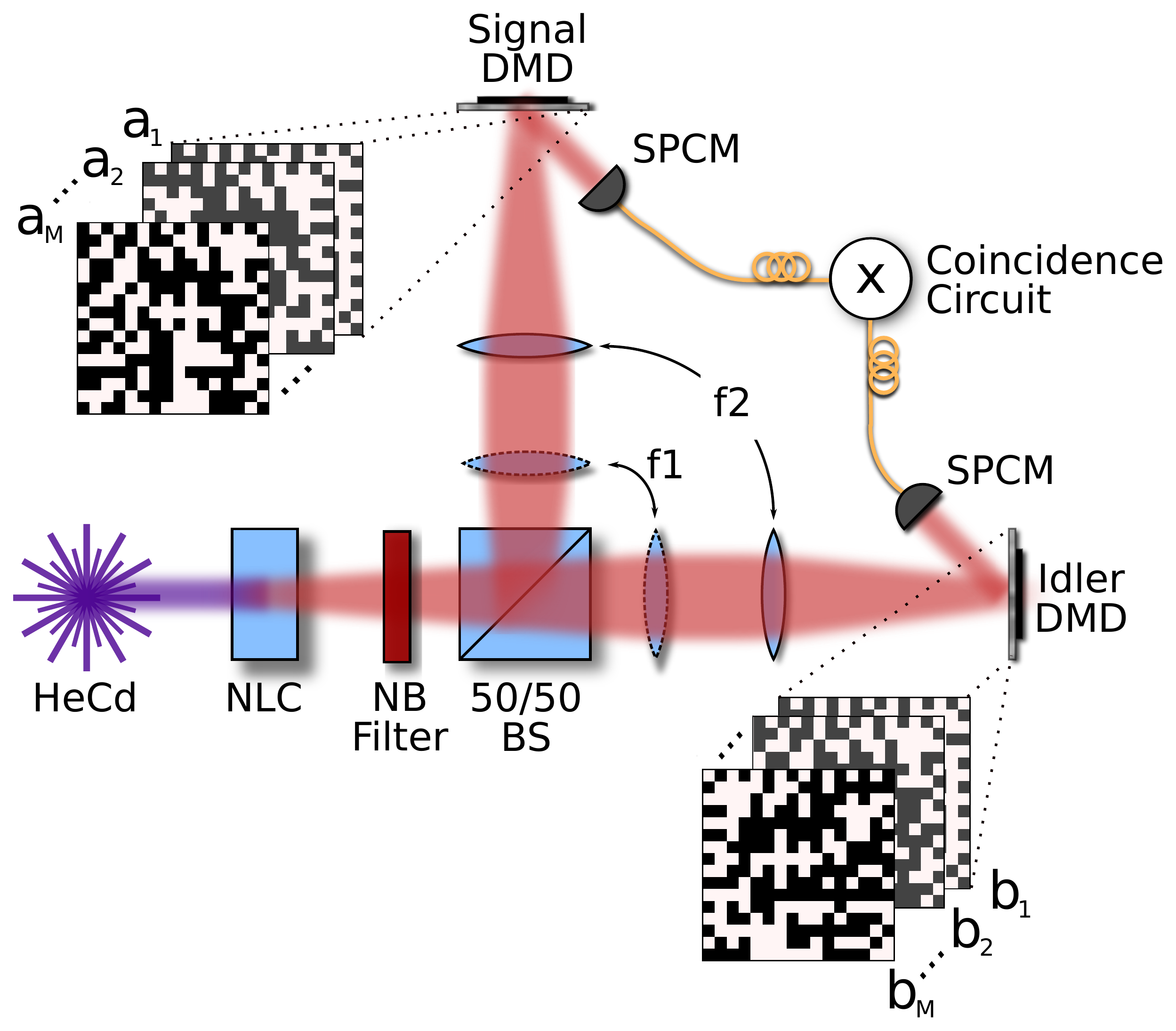}
\caption{Experimental Setup. Photons generated via SPDC pass through a
  narrow-band filter and are split into signal and idler modes by a
  50/50 beamsplitter. For position correlations, lenses $f1 = 125$ mm
  and $f2 = 500$ mm form a 4$f$ imaging system with the crystal and
  DMDs placed in the object and image planes respectively. For
  momentum correlations, f1 is removed and the DMD is placed in the
  focal plane of $f2 = 88.3$ mm. Photons striking DMD ``on'' pixels
  are directed to large area, single-photon counting modules. Photon
  arrivals are then correlated by a coincidence circuit.}
\label{fig:setup}
\end{figure}

\begin{figure*}[!htbp]
  \includegraphics[scale=0.5]{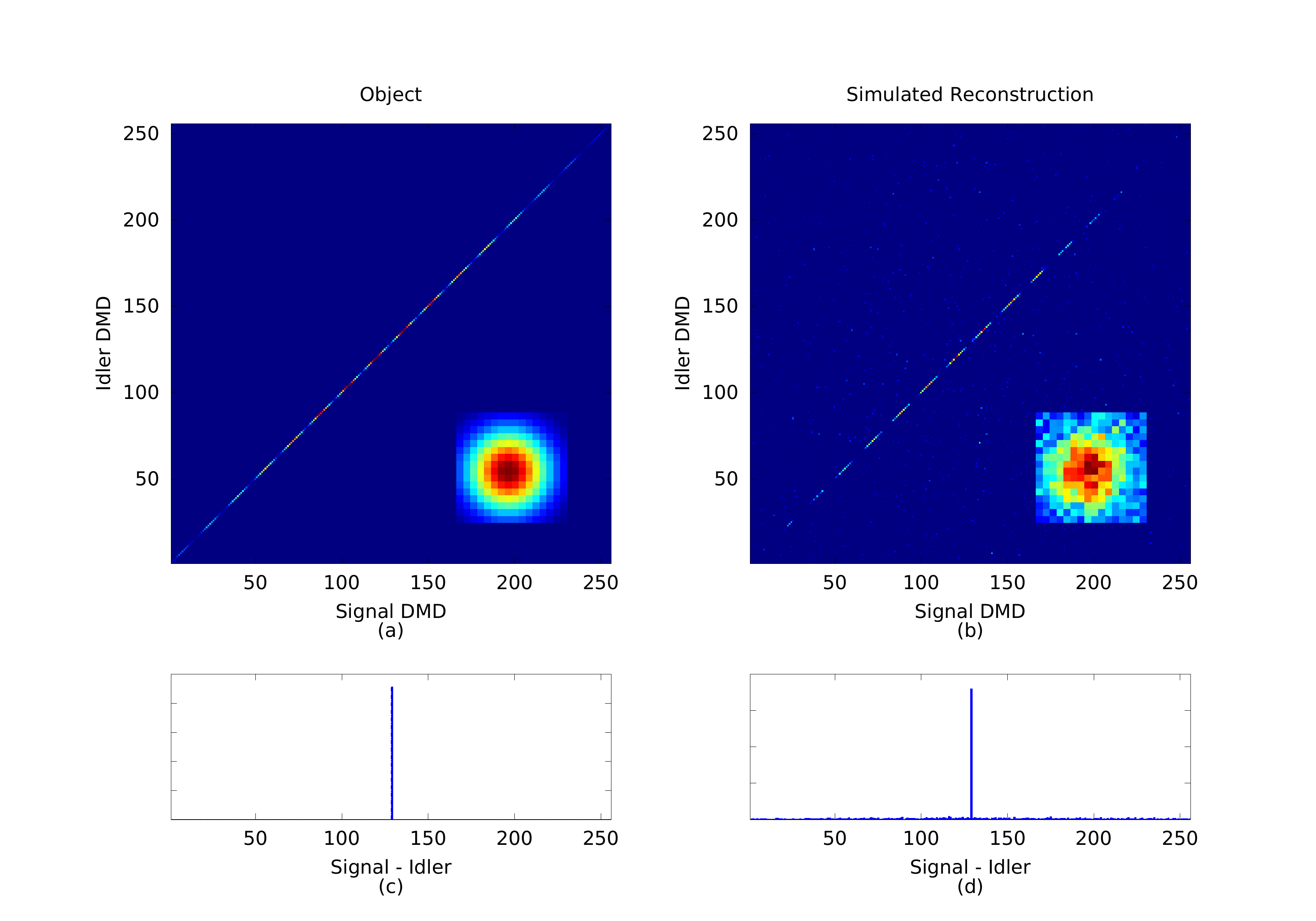}
  \caption{$16\times 16$ Pixel Simulation. The ideal object is given in
    (a). The object was incoherently sampled with $m = 2500$ random
    binary patterns. Poissonian noise corresponding to 5000 photons in
    the field ($\approx 1250$ detected) per measurement was added to
    the measurement vector. The reconstruction is shown in (b), with
    MSE $5\times 10^{-8}$. Plots (c) and (d) integrate along the
    anti-diagonal to show that the reconstruction recovers the
    correlation width $\sigma_c < 1$ pixel with negligible error.}
  \label{fig:simrecon}
\end{figure*}

\begin{figure}[!htbp]
  \includegraphics[scale=1.0]{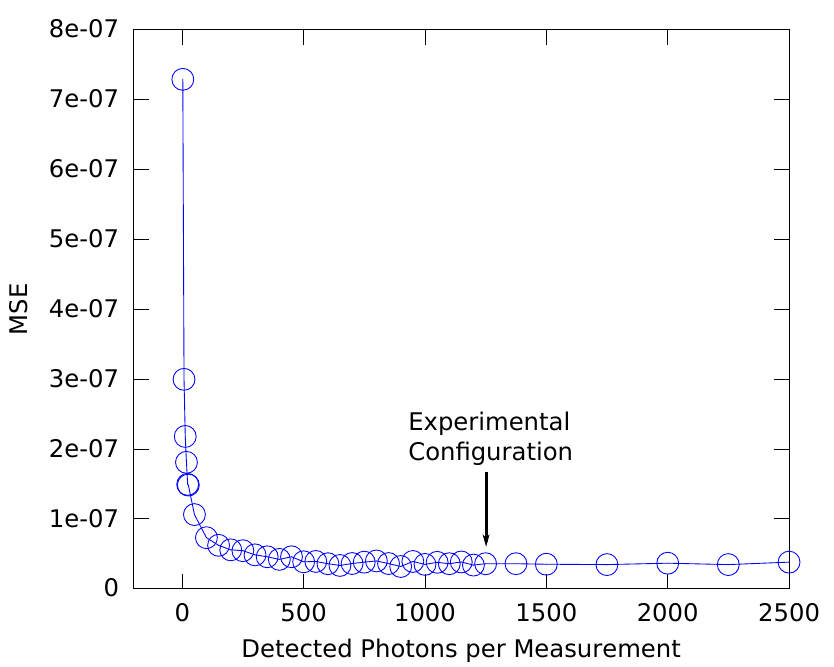};
  \caption{Simulated Mean Squared Error (MSE) versus Photon Flux for
    $n=256$ and $m=2500$. The phase change behavior versus photon
    number can be clearly seen. The experiment used $5000$ total (1250
    detected) photons per measurement to comfortably exceed the phase
    change. The MSE approaches a value $5\times10^{-8}$ corresponding
    to an SNR of roughly $17$.}
  \label{fig:mse}
\end{figure}

\begin{figure*}[!htbp]
  \begin{centering}
    \includegraphics[scale = 0.5 ]{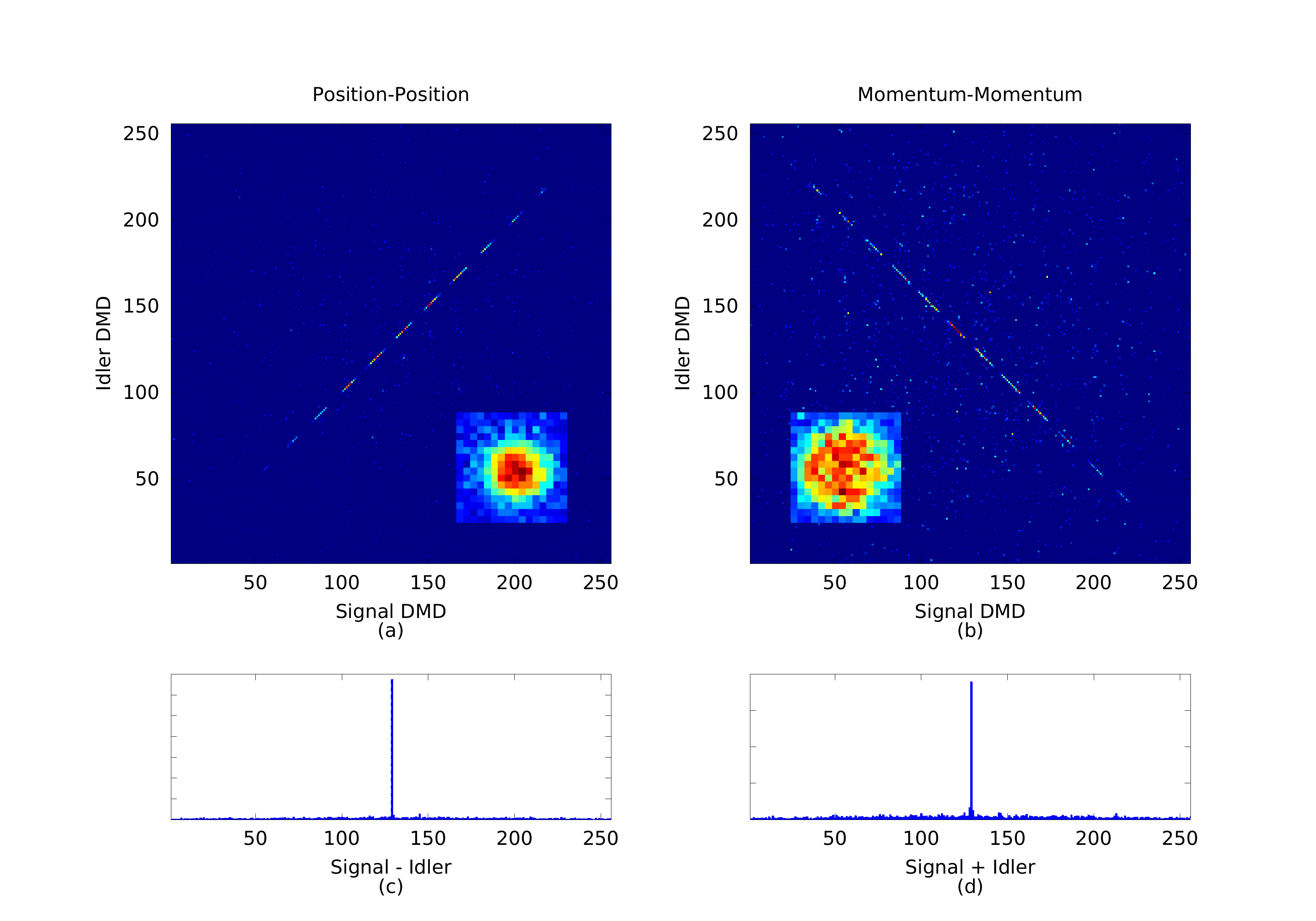}
    \caption{Sample $16\times 16$ Experimental
      Reconstructions. \ref{fig:recon}(a) and \ref{fig:recon}(b) give
      the joint probability distribution for position-position and
      momentum-momentum correlations where DMD pixels are listed in
      column-wise order. 2D marginal distributions for the signal
      photon are inset. \ref{fig:recon}(c) and \ref{fig:recon}(d) show
      correlation widths of only 1 pixel by summing over the signal
      $+$ idler (c) and signal $-$ idler (d) axes. Only 2500 ($3 \%$
      of raster-scanning) measurements were needed with a total
      acquisition time of about 40 minutes.}
    \label{fig:recon}
  \end{centering}
\end{figure*}

\begin{figure}[!htbp]
  \includegraphics[scale=1.0]{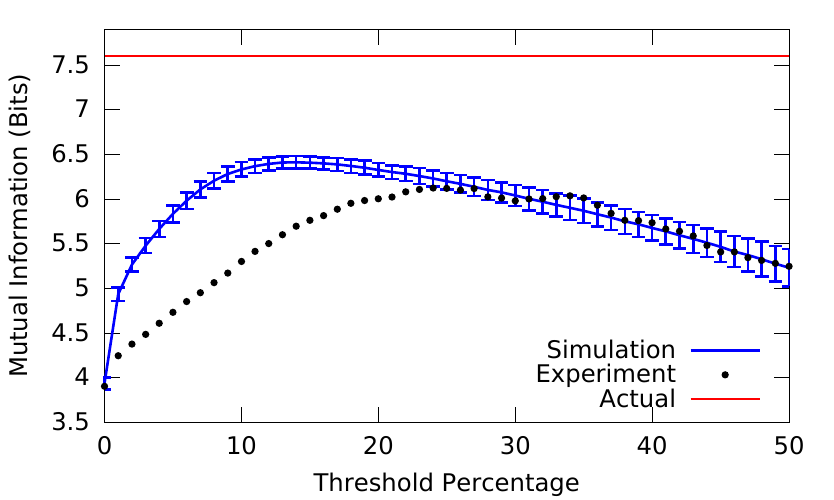};
  \caption{Mutual Information Versus Thresholding. The mutual
    information for reconstruction values above a thresholded
    percentage of the maximum is given for $100$, $n=256$ pixel
    simulations with $m=2500$ measurements and $\Phi = 5000$ photons
    per measurement. The red, constant line gives the true mutual
    information for the simulated object. The black points give the
    $n=256$ far field experimental data for comparison.}
  \label{fig:thresh}
\end{figure}

\begin{figure*}[!htbp]
  \includegraphics[scale=1.0]{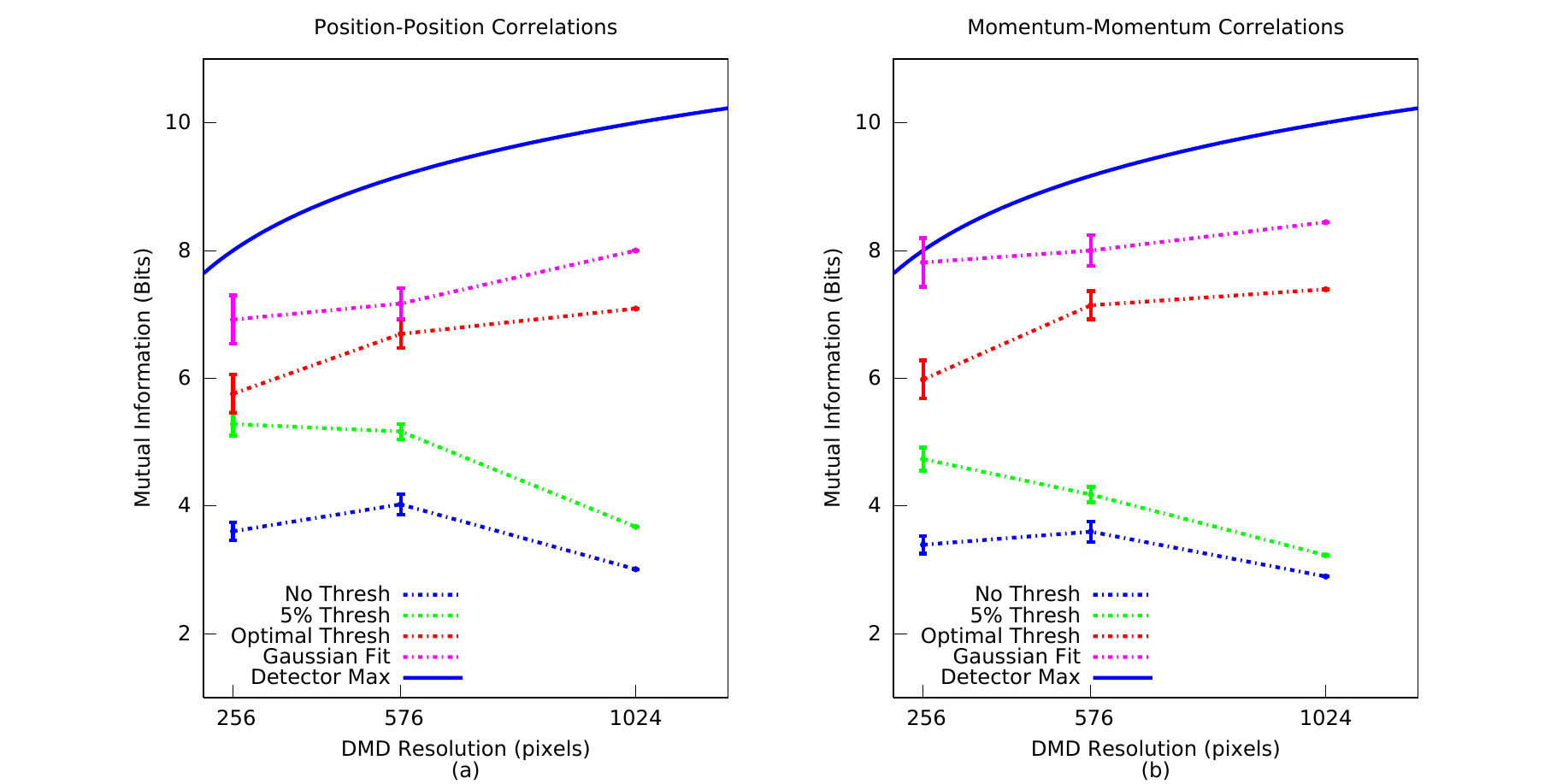}
  \caption{Mutual information between signal and idler photons for
    position-position and momentum-momentum representations are
    presented as a function of detector resolution. Three levels of
    thresholding are shown, as well as a fit to
    Eq. \ref{eq:wavefunction}. Dashed lines are to guide the
    eye. Error bars enclose two standard deviations from the expected
    uncertainty from simulations (not performed for $n=1024$). The
    solid curve represents the maximum possible value for a particular
    detector resolution given perfect correlations and uniform
    marginals.}
  \label{fig:info}
\end{figure*}

\begin{figure}[!htbp]
\includegraphics[scale=1.0]{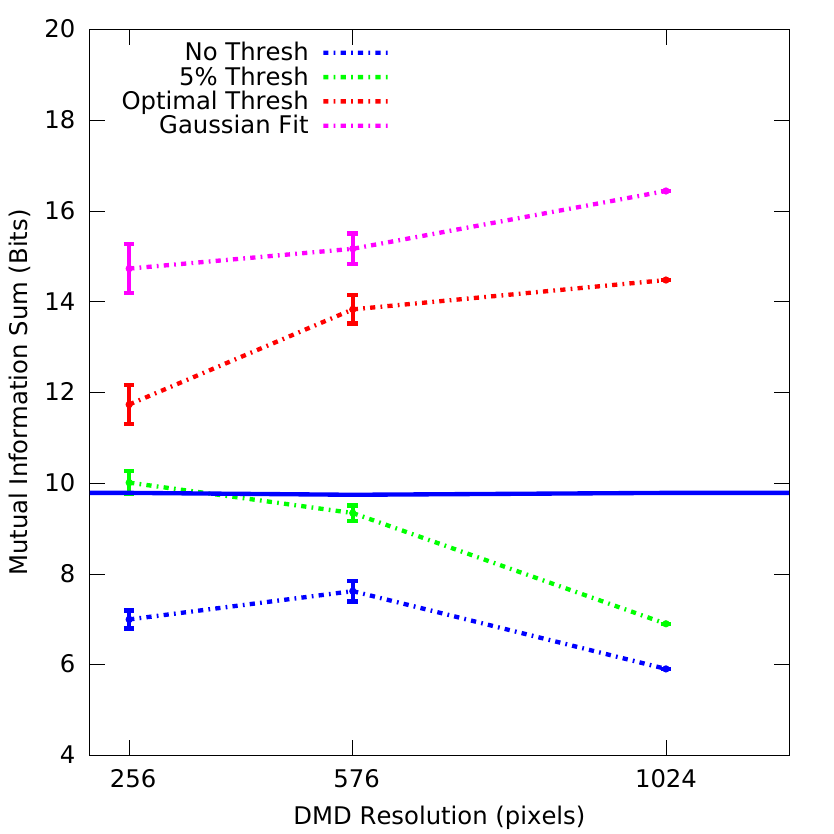}
\caption{The sum of position-position and momentum-momentum mutual
  information is presented as function of detector resolution to
  demonstrate violation of an EPR steering inequality
  (Eq. \ref{eq:steering}). The solid line represents the threshold
  that must be \emph{exceeded} to witness EPR steering. Error bars
  enclose two standard deviations from the expected uncertainty from
  simulations (not performed for $n=1024$). Because simulations
  systematically underestimate the mutual information, values above
  the bound are convincing.}
\label{fig:epr}
\end{figure}

The particular algorithm chosen to solve Eq. \ref{eq:objective} also
plays a role in the reconstruction''s accuracy. These often have
provable performance on ideal signals, but degrade when confronted
with noisy or otherwise non-ideal conditions. In these circumstances,
they have various strengths, including speed, accuracy, and
sensitivity to user selected parameters such as $\tau$ in
Eq. \ref{eq:objective}. For more information on common reconstruction
algorithms, see Refs
\cite{figueiredo:2007,candes:2008,li:2010,gill:2011}.

In practice, the best way to determine accuracy for a particular
signal, sensing matrix, and reconstruction approach remains repeated
simulations or experiments. For our system, we reduce a $n=32\times32$
measurement from a $310$ day raster scan (SNR of 10) to an $8$ hour
compressive acquisition, a thousand-fold improvement.

\section{Experiment}
The experimental apparatus is given in Fig. \ref{fig:setup}. Light
from a $2.8$ mW, $325$ nm HeCd laser was directed to a $1$ mm long
BiBO crystal oriented for type I, collinear SPDC. The generated
daughter photons passed through a $650/13$ nm narrowband filter before
separating into signal and idler modes at a $50/50$ beamsplitter. To
measure position-position correlations, lenses $f1=125$ mm and
$f2=500$ mm imaged the crystal onto signal and idler mode DMDs. For
momentum-momentum correlations, $f1$ was removed and the DMDs were
placed in the focal plane of $f2=88.3$ mm. DMD ``on'' pixels reflected
light to large area, single photon counting modules (SPCM) connected
to a correlating circuit.

To measure $p(u,v)$, a series of $M$ random patterns were placed on
the DMDs to form the sensing matrix $A$. For each set of patterns,
joint detections were counted for acquisition times $t_{aq}$ for a
total measurement time $t = Mt_{aq}$ to make up the measurement vector
$y$. The joint distribution $p(u,v)$ was reconstructed using a
Gradient Projection solver for Eq. \ref{eq:objective} with $\ell_{1}$
regularization, commonly referred to as Basis Pursuit Denoising
\cite{figueiredo:2007}.

We measured at dimensions of $N=256^2$, $N=576^2$, and $N=1024^2$
corresponding to DMD resolutions of $16\times 16$, $24\times 24$, and
$32\times 32$ pixels. The associated measurement numbers $M$ were
$2500$, $10,000$, and $30,000$ so that $M$ is only about
$0.03N$. Acquisition times were $1$ second for position measurements
and $1.5$ seconds for momentum measurements to average $1000$
coincident detections per DMD configuration in all
cases. Additionally, we performed representative simulations at
$16\times 16$ and $24\times 24$ resolutions.

\section{Results}
\subsection{Joint Probability Distribution}

A simulation for measuring position-position correlations at $16\times
16$ DMD resolution is given in Fig. \ref{fig:simrecon}. The object in
\ref{fig:simrecon}(a) is the correlation function of
Eq. \ref{eq:p}. The simulation used $m=2500$ measurements and a photon
flux of $\Phi = 5000$ photons/measurement multiplied by the ideal
$p(u,v)$, conditions representative of the $1$ second experimental
acquisitions. Note that this is the total signal strength before
interacting with the sensing matrix; the mean value of the measurement
vector is $\Phi/4 = 1250$ detected photons. Values of the measurement
vector were Poissonian distributed to simulate the effect of shot
noise.

Fig. \ref{fig:recon}(b) gives the reconstructed correlation function
$p(u,v$) between signal and idler DMD pixels. The sharply defined
diagonal line shows the expected positive correlations between the two
DMDs. DMD pixels are listed in column-major order. The mean squared
error (MSE) for the reconstruction was $5\times10^{-8}$.  The
two-dimensional signal marginal distribution is inset, which provides
an image of the signal beam.  Figs. \ref{fig:simrecon}(c) and
\ref{fig:simrecon}(d) sum the result along the anti-diagonal to show
the correlation width $\sigma_c$. Qualitatively, the reconstruction
closely resembles the original object, faltering only near the edges
of the distribution where the signal falls beneath a noise floor. The
reconstruction recovers $\sigma_c<1$ pixel with negligible error.

To demonstrate the reconstruction accuracy, simulations were performed
for increasing photon flux $\Phi$ with DMD resolution $16\times 16$
and $m=2500$. The MSE versus $\Phi$ is given in Fig. \ref{fig:mse}.
Reconstructions were normalized to the incident flux $\Phi$ for
comparison to the ideal signal. The result shows the rapid phase
change from poor to excellent reconstructions with a MSE converging to
$5\times10^{-8}$ beyond the phase change.

The MSE can be used to roughly estimate the signal-to-noise ratio for
a particular measurement of an average, non-zero element. Assuming
perfect pixel correlations and uniform marginal distributions, the
energy in the signal is distributed over $1/n$ elements. The
signal-to-noise ratio is then $1/n\sqrt{\text{MSE}}$. For $n=256$
pixels and MSE $=5\times10^{-8}$, this yields an approximate SNR of
$17$. For comparison, using Eq. \ref{eq:raster}, a raster scan would
require about four days to achieve a SNR of only $10$. The simulated
CS acquisition time was $42$ minutes for $2500$, $1$ second
measurements.

Sample experimental reconstructions for position-position and
momentum-momentum correlations at $16 \times 16$ pixel DMD resolution
are given in Fig.\ref{fig:recon}. As in the simulations, the
position-position result shows a well defined diagonal line indicating
positive pixel correlations. Conversely, the momentum-momentum result
shows an anti-diagonal line showing the expected
anti-correlations. Figs.  \ref{fig:recon}(c) and \ref{fig:recon}(d)
sum the results along the anti-diagonal (position-position) and
diagonal (momentum-momentum) to reveal an effective correlation width
$\sigma_{ce}$ of a single pixel. Our detection scheme is therefore as
accurate as possible at this resolution and our channel capacity
remains detector limited.

\subsection{Mutual Information in the Channel}
Once $p(u,v)$ is recovered, the channel capacity is given by the
classical mutual information shared between signal and idler DMD
pixels;
\begin{align}
   I =&  -\sum_u p(u)\log p(u) - \sum_v p(v)\log p(v) \nonumber
   \\ & + \sum_{u,v}p(u,v)\log p(u,v), 
\end{align}
where for example,
\begin{equation}
p(u) = \sum_v p(u,v)
\end{equation}
is the signal particle's marginal probability
distribution \cite{dixon:2012}. This entropic analysis is solely
measurement based and does not require reconstructing a wavefunction
or density matrix, a challenging task even for low-dimensional systems
\cite{barnett:1989, walborn:2009, walborn:2011}.

To estimate the uncertainty in the mutual information from shot noise
and the reconstruction process, we performed $100$ simulations at
$n=256$ pixel resolution and $31$ simulations at $n=576$ pixel
resolution. Simulations were not performed at $n=1024$ pixel
resolution due to available computer time. In addition to the results
from the raw reconstruction, thresholding was performed to provide
noise reduction, where all values in the recovered $p(u,v)$ below a
percentage of the maximum value are forced to zero. The simulated
mutual information versus thresholding percentage is given in
Fig. \ref{fig:thresh} for the $n=256$ pixel simulations exemplified by
Fig. \ref{fig:simrecon}. Errorbars enclose one standard deviation from
repeated simulations.

As the threshold increases from zero, the mutual information rises as
a weak, uncorrelated noise floor is removed. An optimal threshold is
quickly reached, beyond which the threshold removes more signal than
noise, reducing the mutual information. Note that the reconstructed
mutual information is systematically lower than the actual mutual
information in the ideal object. This is due to remaining noise and
difficulty in recovering parts of the signal towards the tail of the
distribution.

The $n=256$ far field experimental result is included for comparison
to the simulation. The experiment closely matches the simulation both
for no thresholding and beyond its optimal threshold, but is smaller
in the intermediate region. This is likely due to experimental errors
not included in the simulation. These include slight pixel
misalignment between signal and idler DMDs, optical aberrations,
detector dark noise, stray light, power fluctuations in the laser, and
temperature stability of the nonlinear crystal. Fig. \ref{fig:thresh}
indicates that these experimental difficulties appear to increase the
uncorrelated noise floor rather than significantly affect the
correlated part of the reconstruction.

Although thresholding is a simple post-processing technique, it is
applicable to how the entangled pixels might be used for
communication. If a pair of entangled pixels has a correlated
amplitude near or below the background noise, it will be difficult to
use that particular pair for communication. A communication scheme
would likely only consider pixel pairs above a certain threshold to be
useful. This is related to the technique in photonic quantum
information of subtracting background noise from a measured signal. In
CS, it is also common to perform post-processing or secondary
optimization after maximizing sparsity, such as the debiasing routine
used in Ref. \cite{figueiredo:2007}.

The experimental channel capacity versus DMD resolution for both
position-position and momentum-momentum is given in
Fig. \ref{fig:info} for several levels of thresholding. The optimal
threshold is that which maximizes the mutual information. At $256$ and
$576$ pixel resolutions, optimal thresholds of $20\%$ and $30\%$ were
used for position-position and momentum-momentum distributions
respectively. At $1024$ pixel resolution, noise was more significant,
so the optimal thresholds increased to $30\%$ and $40\%$. Error bars
on $n=256$ and $n=576$ pixels measurements represent the expected
effect of shot noise and reconstruction uncertainty derived from
simulation. These have been conservatively set to include two standard
deviations from the simulated result.

The joint probability distribution was also fit to the double-Gaussian
wavefunction (Eq. \ref{eq:wavefunction}) to find effective widths
$\sigma_{ce}$ and $\sigma_{pe}$. When $\sigma_p \gg \sigma_c$, the
mutual information between particles for Eq. \ref{eq:wavefunction} is
the logarithm of the Federov ratio \cite{federov:2009}
\begin{equation}
  \log \left(\frac{\sigma_p^2}{\sigma_c^2}\right),
  \label{eq:federov}
\end{equation}
where the ratio is squared for two dimensions. While this technically
applies to the continuous wavefunction, and the true $\sigma_c$ is
smaller than a DMD pixel, Eq. \ref{eq:federov} still applies to the
discritized measurement so long as the effective $\sigma_{ce} \gg
\sigma_{pe}$.

Fitting yielded the largest channel capacities with a maximum of $8.4$
bits for momentum-momentum correlations at $1024$ pixel resolution,
equivalent to $337$ independent, identically distributed, entangled
modes. 

Given that fitting more accurately characterizes the system and gives
a larger mutual information, it is reasonable to question the
usefulness the direct mutual information computation. However, the two
approaches suit different purposes. Fitting is useful if one is
particularly interested in the state itself. However, if one intends
to use correlated pixels for some other purpose, such as
communication, the direct calculation is more appropriate. This is
because the correlated pixels on the low intensity, tail of the
distribution will be difficult to use in practice even if their
amplitude can be inferred by fitting.

The solid curve of Fig. \ref{fig:info} gives the maximum possible
mutual information between two, $n$-pixel detectors. Assuming perfect
diagonal or anti-diagonal correlations and uniform marginals, this is
simply $\log(n)$. Because we have Gaussian marginals, we do not expect
to reach this bound, even with $\sigma_{ce} \approx 1$ pixel. By
magnifying and using only the central part of the field, we could
approach this upper limit.

\subsection{Witnessing Entanglement}
Despite not reconstructing a full density matrix, it is still possible
to demonstrate non-classical behavior by comparing position-position
and momentum-momentum correlation measurements directly. This has
traditionally involved fitting the measurements to Eq.
\ref{eq:wavefunction} and analyzing products or sums of conditional
variances \cite{reid:1989,duan:2000,simon:2000}.

We recently presented a more inclusive, entropic steering inequality
for witnessing continuous variable entanglement with discrete
measurements \cite{schneel:2012}, where the sum of the classical
mutual information between position-position and momentum-momentum
correlations is classically bounded. For our system, all classically
correlated measurements must satisfy
\begin{equation}
I_{xs,xi}+I_{ks,ki} \le 2\log \left(\frac{n d_xd_k}{\pi e} \right),
\label{eq:steering}
\end{equation}
where $d_x$ and $d_k$ are the respective widths of DMD pixels in the
position and momentum basis. Note that $nd_xd_k$ is simply the
bandwidth product for the DMD area, and is independent of $n$ if total
area does not change.

The sum of the classical mutual information in conjugate bases for
each detector resolution is given in Fig. \ref{fig:epr}. The solid
blue line provides right hand side of Eq. \ref{eq:steering}, which
must be exceeded to witness EPR steering. Error bars for the $n=256$
and $n=576$ cases are derived from simulation and include two
standard deviations. In all cases, we show EPR steering both with
optimal thresholding and fitting to the double Gaussian wavefunction
(Eq. \ref{eq:wavefunction}). Even at $5\%$ thresholding, there is a
violation for $16 \times 16$ dimensions. Recall that simulations
(Fig. \ref{fig:mse}) systematically under-represented the object
mutual information relative to measurement uncertainty, so
measurement error is highly unlikely to have over-estimated this
sum. For the fitted $32\times32$ dimensional result, we violate the
classical bound by $6.6$ bits.

\section{Conclusion}
In this article, we present a compressive sensing, double-pixel camera
for characterizing the SPCD biphoton state with photon-counting
detectors. This technique is very efficient, improving acquisition
times over raster-scanning by $n^2/\log(n)$ for $n-$pixel
detectors. We image SPDC correlations at up to 1024 dimensions per
detector and measure a detector-limited mutual information of up to
8.4 bits. We also violate an entropic EPR steering bound, indicating
that these correlations are non-classical. More broadly, our results
suggest that compressive sensing can be extremely effective for
analyzing correlations within large dimensional signals
(eg. intensity-intensity correlations). Potential applications range
from verifying security in spectral correlations for energy-time QKD
\cite{ali-khan:2007} to imaging through scattering media
\cite{gong:2011}.

This work was supported by DARPA InPho grant W911NF-10-1-0404.


%

\end{document}